\documentclass[12pt,preprint]{aastex}

\begin{document}

\title{Hierarchical Star Formation in Nearby LEGUS Galaxies}

\author{Debra Meloy Elmegreen\altaffilmark{1},
Bruce G. Elmegreen\altaffilmark{2}, Angela Adamo\altaffilmark{3,4},
Alessandra Aloisi\altaffilmark{5}, Jennifer Andrews\altaffilmark{6}
Francesca Annibali\altaffilmark{7}, Stacey N. Bright\altaffilmark{5},
Daniela Calzetti\altaffilmark{6}, Michele Cignoni\altaffilmark{5},
Aaron S. Evans\altaffilmark{8,9},
John S. Gallagher III\altaffilmark{10}, Dimitrios A. Gouliermis\altaffilmark{11},
Eva K. Grebel\altaffilmark{12}, Deidre A. Hunter\altaffilmark{13}
Kelsey Johnson\altaffilmark{8},
Hwi Kim\altaffilmark{14}, Janice Lee\altaffilmark{5}, Elena Sabbi\altaffilmark{5},
Linda Smith\altaffilmark{15}, David Thilker\altaffilmark{16},
Monica Tosi\altaffilmark{7}, Leonardo Ubeda\altaffilmark{5}}
\altaffiltext{1}{Vassar College, Dept. of Physics and Astronomy, Poughkeepsie, NY 12604}
\altaffiltext{2}{IBM Research Division, T.J. Watson Research Center, Yorktown Hts., NY 10598}
\altaffiltext{3}{Max Planck Institut f\"ur Astronomie, K\"onigstuhl 17, D-69117 Heidelberg, Germany}
\altaffiltext{4}{Department of Astronomy, Oskar Klein Centre, Stockholm University, AlbaNova University Centre, SE-106 91 Stockholm, Sweden}
\altaffiltext{5}{Space Telescope Science Institute, 3700 San Martin Drive, Baltimore, MD 21218, USA}
\altaffiltext{6}{Department of Astronomy, University of Massachusetts, Amherst, MA 01003, USA}
\altaffiltext{7}{INAF-Osservatorio Astronomico di Bologna, Via Ranzani 1, I-40127 Bologna, Italy}
\altaffiltext{8}{Department of Astronomy, University of Virginia, P.O. Box 400325, Charlottesville, VA 22904-4325, USA}
\altaffiltext{9}{National Radio Astronomy Observatory, 520 Edgemont Road, Charlottesville, VA 22903}
\altaffiltext{10}{Department of Astronomy, University of Wisconsin-Madison, WI 53706, USA}
\altaffiltext{11}{Universit\"at Heidelberg, Zentrum f\"ur Astronomie, Institut f\"ur Theoretische Astrophysik, Albert-Ueberle-Str. 2, D-69120 Heidelberg, Germany}
\altaffiltext{12}{Astronomisches Rechen-Institut, Zentrum f\"ur Astronomie der Universit\"at Heidelberg, M\"onchhofstr. 12-14, D-69120 Heidelberg, Germany}
\altaffiltext{13}{Lowell Observatory, 1400 West Mars Hill Road, Flagstaff, Arizona 86001 USA}
\altaffiltext{14}{School of Earth and Space Exploration, Arizona State University, Tempe, AZ 85287, USA}
\altaffiltext{15}{Space Telescope Science Institute and European Space Agency, Baltimore, MD 21218, USA}
\altaffiltext{16}{Department of Physics and Astronomy, Johns Hopkins University, 3701 San Martin Drive, Baltimore, MD 21218, USA}

\begin{abstract}
Hierarchical structure in ultraviolet images of 12 late-type LEGUS galaxies is
studied by determining the numbers and fluxes of nested regions as a function of
size from $\sim1$ to $\sim200$ pc, and the number as a function of flux. Two
starburst dwarfs, NGC 1705 and NGC 5253, have steeper number-size and flux-size
distributions than the others, indicating high fractions of the projected areas
filled with star formation. Nine subregions in 7 galaxies have similarly steep
number-size slopes, even when the whole galaxies have shallower slopes. The
results suggest that hierarchically structured star-forming regions several
hundred parsecs or larger represent common unit structures. Small galaxies
dominated by only a few of these units tend to be starbursts. The
self-similarity of young stellar structures down to parsec scales suggests that
star clusters form in the densest parts of a turbulent medium that also forms
loose stellar groupings on larger scales.  The presence of super star clusters
in two of our starburst dwarfs would follow from the observed structure if cloud
and stellar subregions more readily coalesce when self-gravity in the unit cell
contributes more to the total gravitational potential.
\end{abstract} \keywords{stars: formation --- ISM: structure --- galaxies: ISM ---
galaxies: star clusters: general}

\section{Introduction}

Interstellar turbulence produces hierarchical structure in the gas
\citep{kritsuk13} and in the stars that form from this gas \citep[see review
in][]{e10}, leading to nested young stellar regions with flocculent spiral arms
\citep{e03} and star complexes \citep{efremov95} on kpc scales, OB associations
on 100 pc scales \citep{gouliermis11}, and dispersed and bound clusters on
parsec scales \citep{feit84,gomez93,larson95,scheepmaker09,bastian11}. The bound
clusters themselves appear to be the densest parts of this hierarchy, where the
fraction of the local gas mass that is dense enough to form stars is high, and
so the efficiency of star formation in the region is high too
\citep{e08,parmentier09}.

Hierarchical structure in young stellar regions is widespread and may be
characteristic of all star formation. Still, there is considerable variation in
gravitational binding of the clusters that form \citep{larsen00,maiz01}. The
most massive star-forming regions in the Milky Way are mostly unbound, such as
W43, which spans 300 pc containing $7\times10^6\;M_\odot$ of molecular gas and
the potential to form bound clusters up to $\sim10^5\;M_\odot$ \citep{luong11}.
On the other hand, some starburst galaxies \citep[e.g.,][]{whitmore10},
including dwarf irregular starbursts like NGC 1569 \citep{hunter00} and NGC 1705
\citep{annibali09}, have star-forming regions with about the same total mass,
$10^6\;M_\odot$, but concentrated within tightly bound cores spanning only
several tens of pc. We would like to understand why some regions form bound
clusters and others do not.

The kinematic pressure from stellar motions in a massive cluster is $\sim10^8$
k$_B$, much higher than the average molecular cloud pressure, $10^6$ k$_{\rm B}$
for Boltzmann's constant $k_{\rm B}$ \citep{tan13}. Loose stellar groupings have
lower kinematic pressures than clusters. It seems logical that higher pressures
produce a higher fraction of star formation in the form of bound clusters
\citep{e08}. High pressure correlates with high gas surface density in a galaxy
and therefore with high areal star formation rate \citep{kennicutt98}, possibly
giving the correlation between bound cluster fraction and star formation rate
found by \cite{larsen00}, \cite{goddard10} and \cite{adamo11}. Similarly on
smaller scales, the Orion region has a higher pressure than the Sco-Cen region
and Orion also has a higher clustering fraction \citep{elias09}. The extent to
which high pressures influence cluster boundedness for all masses and at all
levels in the hierarchy is unknown.

A related question is whether there is an upper cutoff in the cluster mass
function. A cutoff of $\sim 10^5-10^6\;M_\odot$ was suggested for some spiral
galaxies by \cite{gieles06}, \cite{bastian08}, and \cite{larsen09}. Does the
starburst NGC 1705 mentioned above have a normal cluster mass function but a
higher mass cutoff, allowing a structure with $10^6\;M_\odot$ to become
gravitationally bound? Do starburst galaxies in general have higher cutoffs, or
no cutoffs as suggested for the Antenna galaxy by \cite{whitmore10}?

The formation of super star clusters (SSC) in dwarf galaxies like NGC 1705 is
also important to understand because metal-poor globular clusters probably
formed in dwarf-like galaxies in the early universe
\citep{chies11,elmegreen12,leaman13}. Such a formation site is suggested from
the mass-metallicity relation of galaxies as a function of redshift
\citep{mannucci09}. Perhaps SSCs in small galaxies reach high pressures during
dwarf-dwarf galaxy mergers \citep{bekki08}, or because of the ram pressure from
accreting gas streams, as appears to be the case in NGC 1569 \citep{johnson10}
and NGC 5253 \citep{lopez12}. These perturbations would be large-scale sources
of turbulence, as opposed to stellar feedback, which is a small-scale source.
The scale for turbulent energy injection may be evident from kinks or turn-overs
in the scaling functions for turbulent motions and their resulting structures
\citep{padoan09}.

These questions about the origin and boundedness of stellar groupings, cluster
mass limits, and energy sources for high pressures and turbulence can be
addressed with galaxies selected from the LEGUS survey \citep{calzetti14}. Here
we investigate multi-scale structure of star formation in 12 galaxies. We have
shown previously that the distribution function of region size in a
hierarchically structured region is a power law with a slope that is consistent
with an ISM partitioned by Kolomogorov-like turbulence and viewed in projection
through a galaxy disk \citep{elmegreen06}. The purpose of this study is to see
if the distribution functions for size and luminosity differ for starburst and
normal systems.

\cite{sanchez08} studied the fractal dimension of HII region positions in dwarf
irregular and spiral galaxies, finding that the brightest HII regions in any one
galaxy have smaller fractal dimensions than the faintest HII regions (i.e., the
brighter ones are more clumped together), and that in general the HII region
population in brighter galaxies has a slightly smaller fractal dimension than it
does in fainter galaxies (more clumpy).  They did not consider starbursts,
however.

\cite{parodi03} and \cite{odekon06} determined the correlation dimension
$dN(r)/dr$ for cumulative number of emission points $N$ as a function of
distance $r$ from the centers of star-forming regions in dwarf Irregular
galaxies. They found that the dimension increases for the brighter dwarfs --
meaning that star formation is more area-filling, less strongly sub-clustered,
and less porous for the brighter dwarfs. A similar variation of power spectrum
slope for H$\alpha$ was found by \cite{willett05} in dwarf galaxies, where the
power spectrum slope ranged from the Kolmogorov value characteristic of
turbulence to shallower values as the filling factor of the H$\alpha$ decreased.
Evidently there is a characteristic power-law structure inside all of these star
forming regions, and a dilution of this structure in the whole galaxy depending
on the star formation filling factor.

We find a similar result here, that individual star-forming regions have steep
number-versus-size relations in the NUV, and that the starburst dwarfs have
similarly steep relations throughout their disks because of a dominance of these
structures. The results of this study are in Section 3 and possible implications
are in Section 4.

\section{Observations}

The Legacy Extragalactic UV Survey (LEGUS) is a Hubble Space Telescope Cycle 21
imaging survey in NUV, U, B, V, and I of 50 nearby galaxies with WFC3/UVIS
\citep{calzetti14}. The survey is designed to include galaxies spanning
different Hubble types. The pipeline data reduction is described by
\cite{calzetti14}. In this study, we select 12 galaxies observed at F336W and
F275W in order to examine the distribution of hierarchical structure in the
youngest stars. Composite color images (F275W, F336W, and F435W or F438W) are
shown for nine of the galaxies in Figure 1, while Hubble types and distances are
in Table 1. Four of the galaxies are spirals and the rest are dwarf irregulars,
with two having starburst characteristics and super star clusters, NGC 1705
\citep{annibali09} and NGC 5253 \citep{west13}. Other galaxies in LEGUS are not
included because they are either too highly inclined, incompletely sampled, or
not observed yet.

\section{Data Analysis and Results}

In order to examine the number distribution function of star-forming regions as
a function of size, the F336W and F275W images of each galaxy were smoothed with
the Gaussian function $\it{gauss}$ in IRAF using $\sigma$ values of 2, 4, 8, 16,
32, 64, and 128 pixels (the average FWHM of stars was measured to be
$1.88\pm0.11$ pixels on the F336W image). A sample result is shown in Figure 2.
The source extraction program SExtractor \citep{bertin96} was used to make
catalogs and source images from each blurred image. Different detection
thresholds and minimum area thresholds were tried until realistic-looking source
images were obtained. The fits used a minimum area of 10 pixels, a detection
threshold of $10\sigma$, a local background mesh 64 pixels wide, and a
background filter 3 pixels wide.

The top panels of Figure 3 show the number of sources with a size greater than
the abscissa values versus these sizes for structures in the F336W (left) and
F275W (right) images. The slope in these plots is the projected fractal
dimension of the star formation structure. The two filters give essentially the
same results so there are no strong age effects. The starburst galaxies tend to
have steeper slopes than the spiral and non-burst dwarfs, which means that the
starbursts are more area-filling with lots of small regions inside and around
the large regions. Recall that the dimension approaches the value of 2 as the
projected image becomes totally covered. The slopes do not differ significantly
between the spirals, which are dusty, and the non-starburst dwarfs, which have
less dust, suggesting that extinction is not significant. Neither do the slopes
differ because of the presence of spiral arms, because the largest scales
considered here ($\sim200$ pc) are only comparable to the arm thicknesses and
not to the arms' elongated shapes.

Linear least-squares fits to the correlations discussed here are listed in Table
1. The number-size relation just discussed is fitted by the expression $\log N =
A_{\rm NS} + B_{\rm NS}\log S $ with slope coefficient $B$ in the table and
subscript ``NS'' meaning ``number-size''. Others have a similar notation. In all
cases, the fits are based on the smallest five scales where the correlations are
most like power laws.

The middle left panel in Figure 3 shows the total flux of all the SExtractor
selected regions as a function of size. What is plotted on the ordinate is
\begin{equation}
\log_{10}F=-0.4M_{\rm AB}=\log_{10}C-0.4\times24.5377+2\log_{\rm 10}(10^5D)
\end{equation}
for absolute specific flux $F$, absolute magnitude $M_{\rm AB}$, counts $C$,
zero-point 24.5377 in the case of F336W, and distance $D$ in Mpc. The figure
shows that the total flux decreases slightly with increasing size (slope $B_{\rm
TS}$ in Table 1), which means that most of the selected regions are contained in
one identified structure or another on all scales, except for the small and
faint regions, which drop out successively as the blur size gets larger. The
ratio of this total flux to the number of sources is the average flux per
source; this is shown in the middle-right panel as a function of size (slope
$B_{\rm FS}$). The starbursts again differ from the non-bursting galaxies
because they have a steeper slope in average flux versus size.  This corresponds
to our impression that the starbursts have brighter regions on large scales than
non-starburst galaxies. This brightening occurs systematically for all scales
and not just suddenly at the largest. Because the total flux is nearly invariant
with size for small sizes ($B_{\rm TS}\sim0$), the average flux-size correlation
is approximately the inverse of the number-size correlation ($B_{\rm FS}=B_{\rm
TS}-B_{\rm NS}\sim-B_{\rm NS}$).

The lower left panel shows the flux distribution function (``the luminosity
function,'' slope $B_{\rm NF}$), which replots the ordinate of the number-size
relation versus the ordinate of the average-flux-size relation (the twist at the
bottom of each plot is from the drop in the total flux at large scales, which is
from the loss of faint and small features). If all of the flux in these
structures were present at all scales, then the total flux would be constant
with size, $B_{\rm TS}=0$, and the slope of the flux distribution function would
be $B_{\rm NS}/B_{\rm FS}=B_{\rm NS}/(B_{\rm TS}-B_{\rm NS})=-1$ (for log
intervals) independent of the fractal dimension (which is $-B_{\rm NS}$). Not
all of the flux is present on all scales however ($B_{\rm TS}<0$), because the
smaller and fainter sources that are outliers of the bigger and brighter sources
drop below the $10\sigma$ threshold for inclusion as the Gaussian blur size
increases. NGC 1705 has a flux distribution function slope that is shallower by
$4\sigma$ compared to the others, and also a maximum flux that is nearly an
order of magnitude larger than for the others, reflecting the presence of the
SSC.

To assess how much of the star formation lies outside of the hierarchy, masks
were made on one scale, e.g., the 32-pixel blur, and then the regions on a
factor-of-two smaller scale that are inside and outside the masks were
determined. Figure 4 shows the inner and outer regions of size 16 pixels (i.e.,
compared to the 32-pixel mask) for NGC 5477.

The lower right panel of Figure 3 shows the outlier fraction more
systematically, plotting the luminosity fractions of regions on a scale of N
pixels that are outside the regions having a scale of N+1 pixels, versus the
scale of N pixels.  All of the galaxies have an increasing outlier fraction with
size (slope $B_{\rm OS}$ in Table 1) except for NGC 1705 and UGC 695, which have
similarly rising fractions for small size and then a drop to zero fraction (the
drop begins at the large dot in the figure). Such a drop indicates a
concentration of essentially all of the bright star formation in one large
region, as is also evident from Figure 1.

Power law slopes for the number-size relation were determined in nine
sub-regions of seven galaxies out to typically 8- or 16-pixel blurs, depending
on the region size. The galaxies were NGC 2500, IC 4247, NGC 5253, NGC 5477 (2
regions), NGC 7793, NGC 3738, and IC 559 (2 regions). The slopes were steep for
all sub-regions, averaging $B_{\rm NS}=1.97\pm0.29$.

\section{Discussion and Conclusions}
\label{sect:conc}

The UV images in this survey show two distinct morphologies. One is
characteristic of the large spiral galaxies and low surface brightness dwarfs
where there is patchy and distributed star formation and low emission between
the patches. There is no obvious hierarchical structure among the different
patches, which seem to be independent or strung out along spiral arms, but there
is hierarchical structure inside of them, to the extent that it can be resolved
(e.g., NGC 7793). The other morphology is characteristic of starburst dwarfs or
HII galaxies where the image is dominated by one or two patches of star
formation, which seem large relative to the size of the galaxy. These patches
are well resolved and clearly hierarchical inside. We identify these ultraviolet
patches with giant star complexes such as those studied by \cite{efremov95}.

The hierarchical structure observed by the number-size distribution or the
flux-size distribution is approximately scale free up to the largest scale, as
shown by the good power-law fits. The corresponding fractal dimension is large
for the individual complexes too, which means a steep number-size slope
approaching the limit of 2 for a completely filled and nested region. The
fractal dimension is almost this large for the whole galaxies that are dominated
by one or two complexes (NGC 1705, NGC 5253, UGC 695). Galaxies of the first
morphological type have small fractal dimensions (shallow slopes).

The galaxies dominated by single large complexes also tend to have most of their
smaller regions inside their larger regions, which means that the fractional
luminosity from outliers goes to zero on large scales. In the other galaxies,
this fraction monotonically increases with scale because the complexes are
spread out and get lost with increased blurring as isolated regions (outliers)
rather than as embedded regions.

The power-law structure of star-forming regions in these galaxies is consistent
with the standard model where star formation is regulated by turbulent
processes, such as gas compressions that form successively smaller clouds inside
and around larger clouds \citep[``turbulent fragmentation,''][]{vazquez09}. Such
processes form a similar hierarchy of young stars, with a likely secondary
correlation for star age, making larger regions older in proportion to the
turbulent crossing time \citep{efremov98,fuente09}. The hierarchy has an upper
limit in size beyond which separate regions form independently. This is
consistent with the observation that the 2-point correlation for stars and
clusters decreases as a power law with increasing scale up to about one kpc
\citep{scheepmaker09,bastian11}.

The starbursts in our sample also have SSCs, especially NGC 1705 and NGC 5253. A
high projected density of hierarchical star formation should play a role in the
formation of these clusters because smaller stellar groupings more readily
coagulate and attract each other in a crowded environment, especially in
low-mass galaxies where the binding energy  in the star-forming cloud is a large
fraction of the gravitational potential in the disk at that location. Moreover,
because these structures are power laws, such coagulation should happen all
throughout the cluster mass range, preserving the cluster mass distribution
function. It should affect primarily the largest cluster mass that can form,
which should increase in such a region.

\cite{minniti04} suggest a coagulation origin for a super star cluster in NGC
5128. This interpretation is also consistent with the finding by
\cite{annibali09} that the stars $10-15$ Myr old in NGC 1705 are closer to the
(coeval) SSC than the younger stars ($<5$ Myr), and that there are many other
smaller clusters nearby. In galaxies with more dispersed star formation, the
only remnants of this hierarchical process could be cluster pairs
\citep{dieball02,fuente09b}.

A shift in the correlated properties of young stars around the star-forming
region NGC 346 in the Small Magellanic Cloud, from one that is fractal on large
scales to one that is centrally concentrated with a power law density profile in
the core region, suggests an analogous change in gas density structure when
self-gravity becomes important in a turbulent medium \citep{gouliermis14}.

In conclusion, star formation observed in ultraviolet images with HST shows
hierarchical structure from scales of a few hundred parsecs down to the parsec
scale of individual bound clusters. The clusters therefore appear to form in the
densest parts of a self-gravitating cloud complex that is structured by
turbulence.  Starburst dwarfs tend to have most of their ultraviolet structure
in this form because they have one or two dominant young star complexes that are
each hierarchical inside. Spiral galaxies and low surface brightness dwarfs have
more uniformly dispersed complexes. The presence of dense hierarchical structure
in a galaxy-dominant star complex would seem to favor an increase in the largest
mass cluster than can form without changing the power law slope of the mass
function for the lower mass clusters. This may be the origin of the
Schechter-type mass function that has been observed for clusters, and it may
also explain the apparent variations in the cutoff mass as a function of
environment.

\acknowledgments Based on observations made with the NASA/ESA Hubble Space
Telescope, obtained at the Space Telescope Science Institute, which is operated
by the Association of Universities for Research in Astronomy, Inc., under NASA
contract NAS 5-26555. These observations are associated with program \#13364
(LEGUS), including grants HST-GO-13364.15-A (DME) and HST-GO-13364.14-A (BGE).
This research has made use of the NASA/IPAC Extragalactic Database (NED) which
is operated by the Jet Propulsion Laboratory, California Institution of
Technology under contract with NASA. DAG kindly acknowledges financial support
by the German Research Foundation through grant GO 1659/3-1.

\clearpage

\begin{deluxetable}{lcccccccc}
\tabletypesize{\scriptsize} \tablecolumns{11} \tablewidth{0pt}
\tablecaption{Linear Fits to Correlations}
\tablehead{
\colhead{Galaxy}&
\colhead{Type}&
\colhead{D (Mpc)\tablenotemark{a}}&
\colhead{$B_{\rm NS}$} &
\colhead{$B_{\rm NS,275}$} &
\colhead{$B_{\rm TS}$} &
\colhead{$B_{\rm FS}$} &
\colhead{$B_{\rm NF}$} &
\colhead{$B_{\rm OS}$}
}
\startdata
NGC 1566 & SABbc & 13.20 & $-1.34\pm 0.05$&$-1.30\pm 0.04$&$-0.33\pm 0.05$&$ 1.00\pm 0.03$&$-1.33\pm 0.05$&$ 0.75\pm 0.12$\\
NGC 1705 & SA0pec [Irr] &  5.10 & $-1.86\pm 0.10$&$-1.89\pm 0.30$&$-0.16\pm 0.01$&$ 1.70\pm 0.10$&$-1.09\pm 0.01$&$ 0.10\pm 0.28$\\
NGC 2500 & SBd & 10.10 & $-1.17\pm 0.06$&$-1.18\pm 0.05$&$-0.31\pm 0.07$&$ 0.85\pm 0.03$&$-1.36\pm 0.09$&$ 0.86\pm 0.09$\\
NGC 3738 & Im &  4.90 & $-1.39\pm 0.06$&$-1.39\pm 0.05$&$-0.60\pm 0.05$&$ 0.80\pm 0.02$&$-1.75\pm 0.06$&$ 0.92\pm 0.09$\\
NGC 5253 & Im pec &  3.15 & $-1.51\pm 0.08$&$-1.52\pm 0.14$&$-0.49\pm 0.06$&$ 1.03\pm 0.05$&$-1.47\pm 0.06$&$ 1.00\pm 0.06$\\
NGC 5477 & SAm &  6.40 & $-0.98\pm 0.06$&$-1.14\pm 0.06$&$-0.17\pm 0.04$&$ 0.81\pm 0.05$&$-1.21\pm 0.06$&$ 0.36\pm 0.15$\\
NGC 7793 &  SAd &  3.44 & $-1.62\pm 0.08$&$-1.62\pm 0.09$&$-0.41\pm 0.07$&$ 1.21\pm 0.05$&$-1.34\pm 0.06$&$ 0.42\pm 0.11$\\
IC 4247 & S? [Irr] &  5.11 & $-1.14\pm 0.04$&$-1.17\pm 0.04$&$-0.40\pm 0.06$&$0.75\pm 0.02$&$-1.53\pm 0.09$&$ 0.63\pm 0.06$\\
IC 559 &  Sc [Irr]  &  5.30 &$-1.12\pm 0.14$&$-1.13\pm 0.08$&$-0.39\pm 0.06$&$ 0.74\pm 0.14$&$-1.47\pm 0.13$&$ 1.16\pm 0.30$\\
ESO486-G021 &  S? [Irr] &  9.50 & $-1.47\pm 0.08$&$-1.32\pm 0.09$&$-0.45\pm 0.10$&$ 1.02\pm 0.03$&$-1.43\pm 0.11$&$ 0.72\pm 0.06$\\
UGC 695 & S? [Irr] & 10.90 & $-1.83\pm 0.15$&$-1.70\pm 0.10$&$-0.43\pm 0.04$&$ 1.40\pm 0.12$&$-1.30\pm 0.02$&$ 1.09\pm 0.16$\\
UGC 7408 & IAm &  6.70 & $-0.76\pm 0.12$&$-0.92\pm 0.09$&$-0.11\pm 0.12$&$ 0.66\pm 0.02$&$-1.17\pm 0.18$&$ 1.33\pm 0.23$
\enddata
\tablenotetext{a}{Hubble types are from the NASA/IPAC Extragalactic Database
(http://ned.ipac.caltech.edu); brackets indicate our revised classifications based on the high
resolution images. Distances are from Calzetti et al. (2014) assuming a Hubble constant of 70 km
s$^{-1}$ Mpc$^{-1}$.} \label{results1}
\end{deluxetable}

\clearpage
\begin{figure}\epsscale{.9}
\includegraphics[width=5.5in]{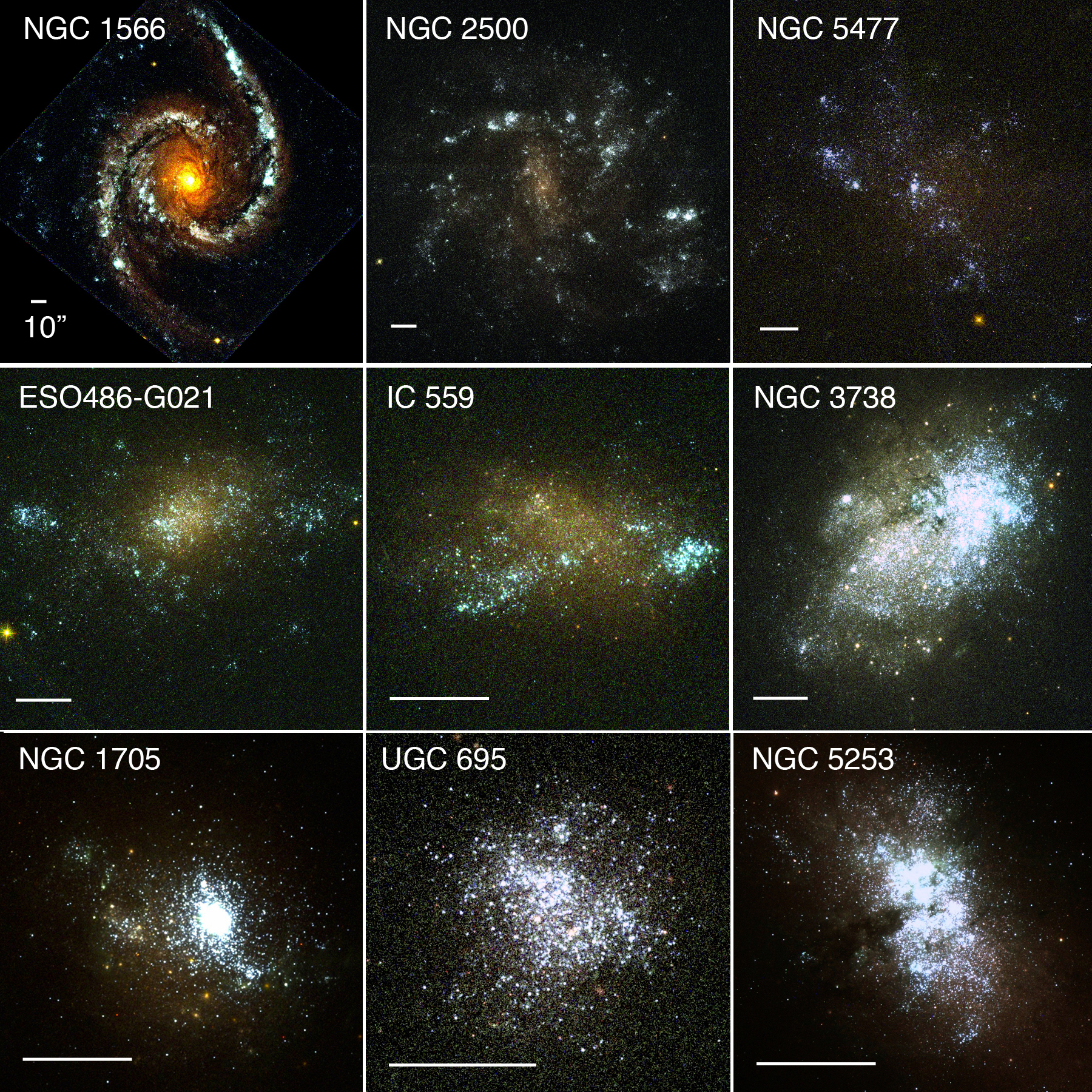}
\caption{HST WFC3/UVIS images for 9 of the 12 galaxies
from the LEGUS survey. Color composites are F275W for B,
F336W for G and F438W for R, all from WFC3, except for NGC 5253 which uses F435W from the ACS.
The scale bar is 10 arcsec.}
\label{fig1}\end{figure}

\clearpage
\begin{figure}\epsscale{.9}
\includegraphics[width=5.5in]{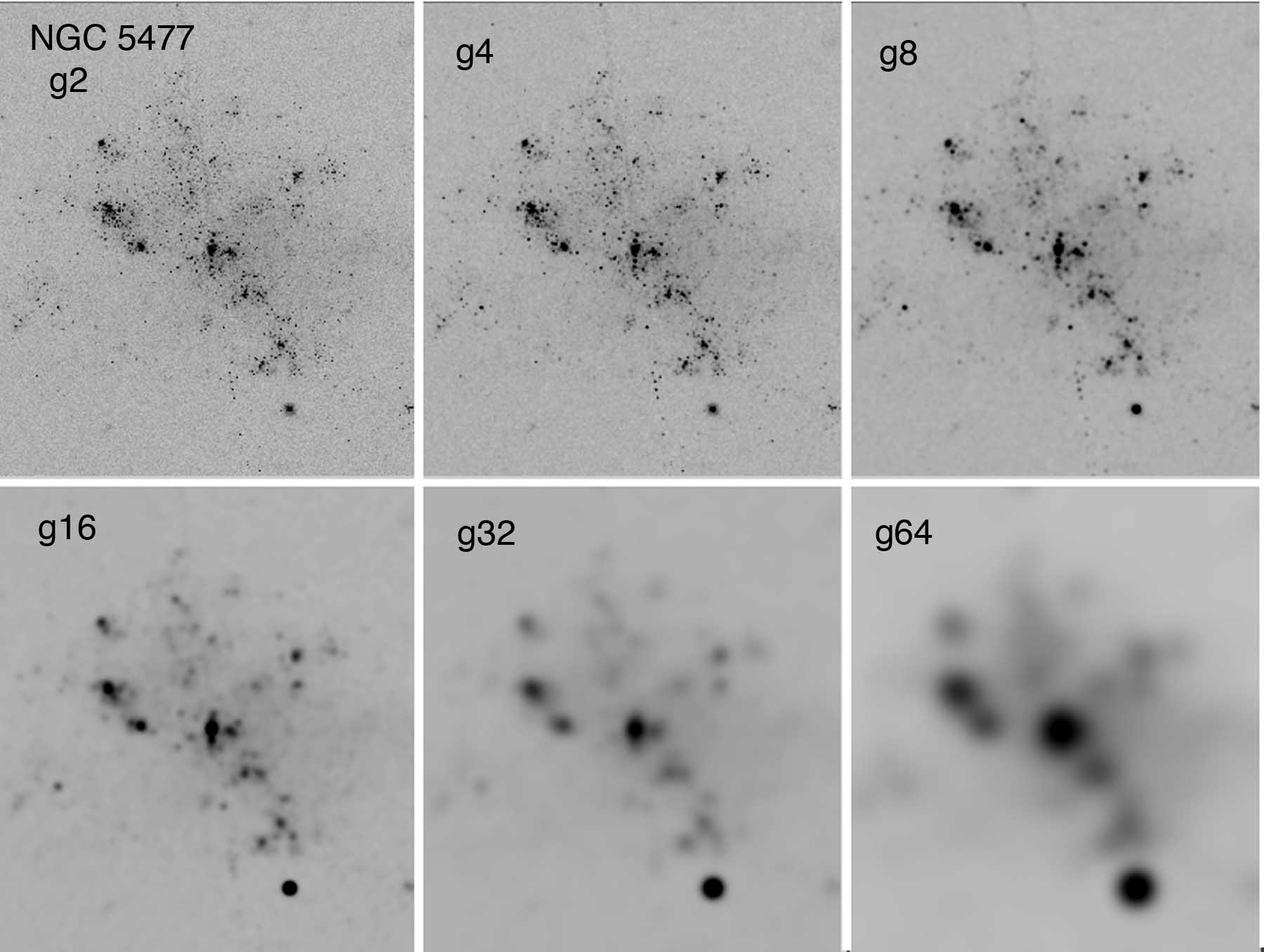}
\caption{Smoothed images of NGC 5477 with Gaussian blurs of 2 pixels, 4, 8, 16, 32, 64.} \label{fig2}\end{figure}

\clearpage

\begin{figure}
\includegraphics[width=4.5in]{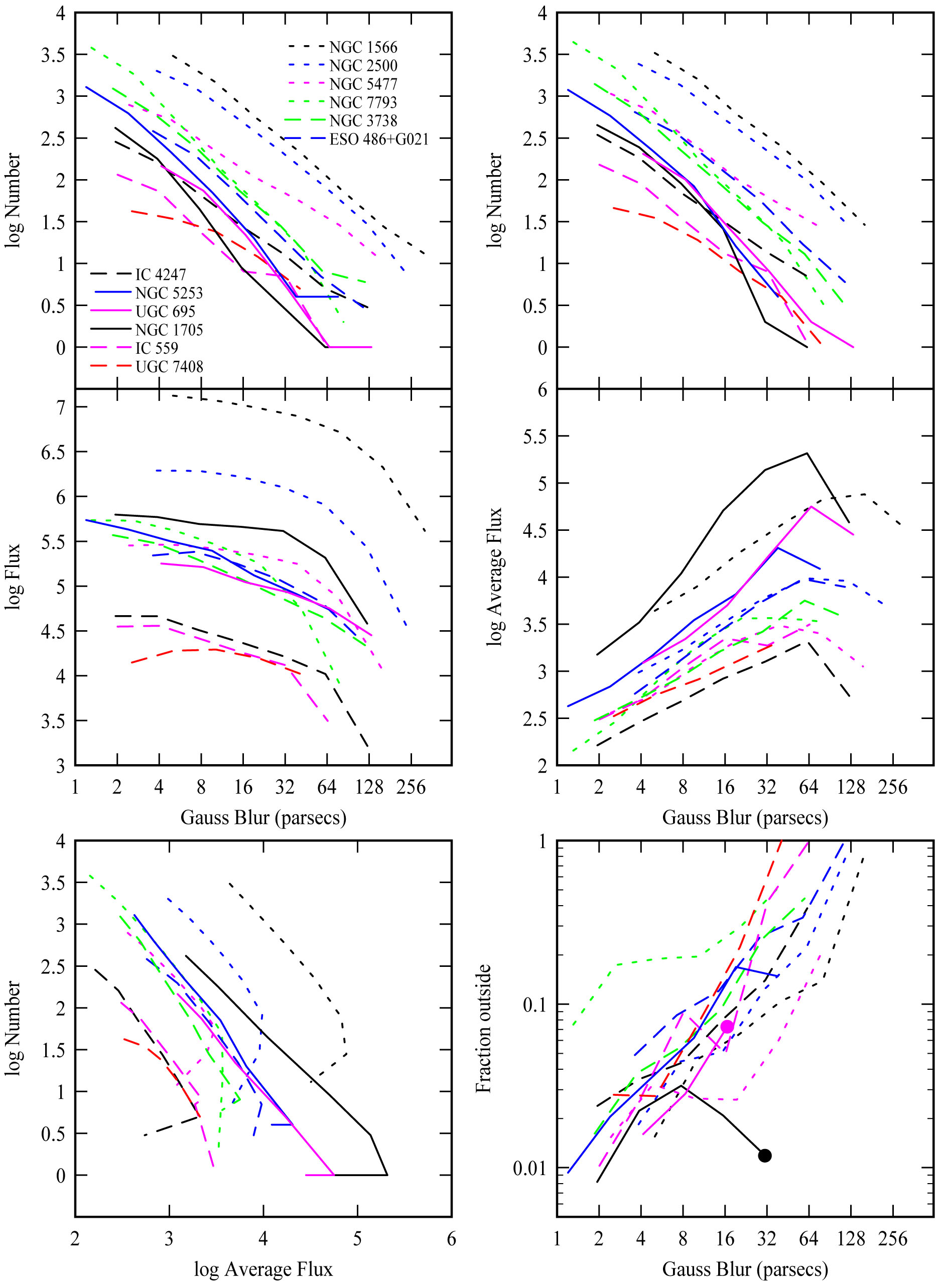}
\caption{Size and flux distribution functions for star-forming regions found by
SExtractor.
Top left: the number of regions larger than size $S$ (in parsecs) versus $S$, from the
F336W images. The galaxies corresponding to each line type are indicated; line types
are roughly divided into spirals (dotted), dwarfs (dashed) and starbursts (lines).
Top right: cumulative number versus size from the F275W images. Middle left:
Total flux at F336W in all SExtractor-selected regions larger than $S$ versus $S$.
Middle right: The ratio of the total flux at F336W to the number of regions larger than $S$
versus $S$; this is the average F336W flux per region. Bottom left: the number of regions
versus their average F336W flux. Bottom right: the fraction of the F336W flux in SExtractor-selected regions
on the plotted scale $S$ that are outside of the SExtractor-selected regions on the
next-larger scale, $2S$.
}\end{figure}

\clearpage
\begin{figure}\epsscale{.9}
\includegraphics[width=5.5in]{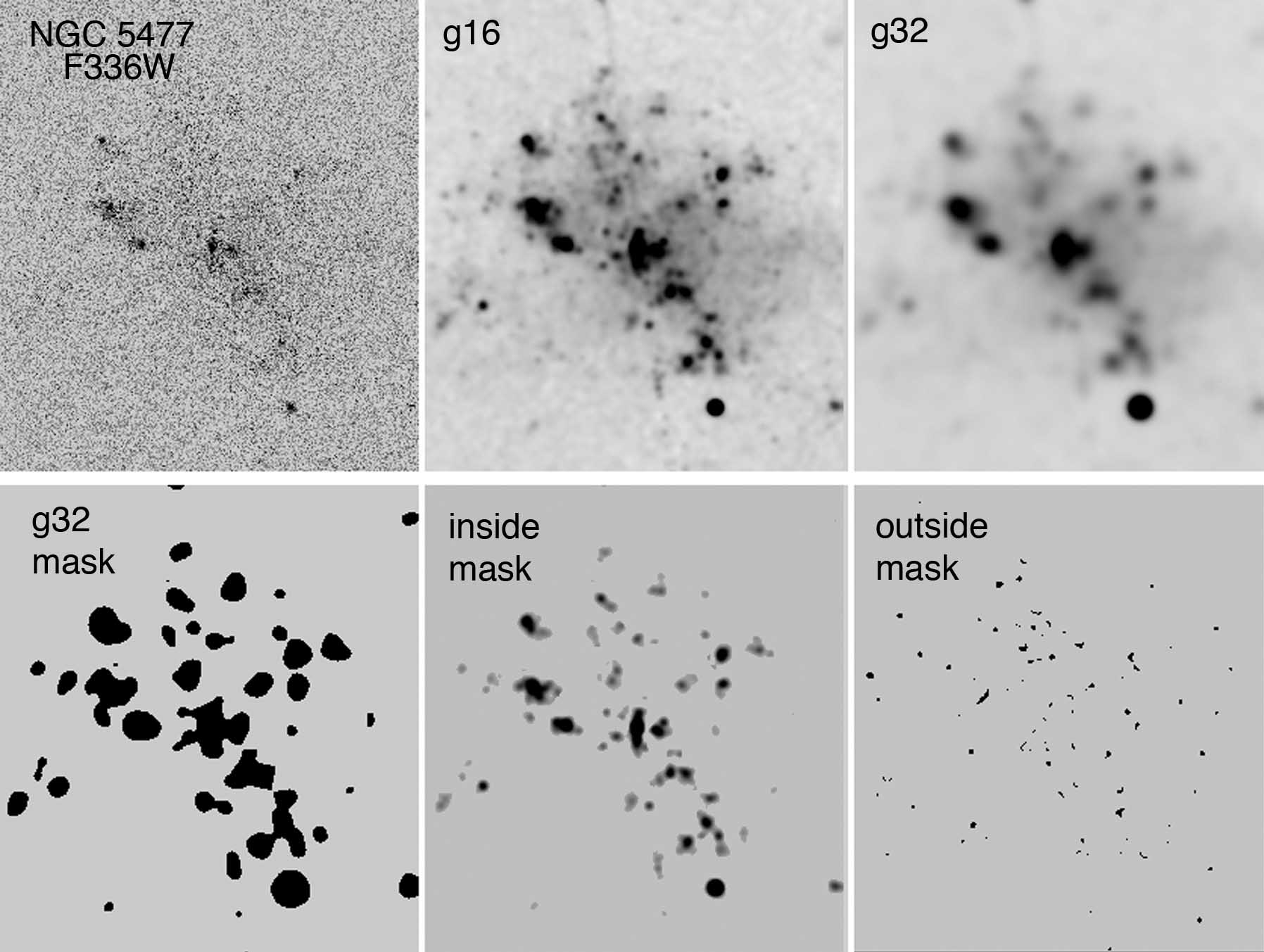}
\caption{NGC 5477 with Gauss blurs of 16 and 32 pixels (left to right, top), the
mask made from the 32 pixel blurred image (lower left), and the g16 sources inside
and outside the mask boundaries.} \label{fig4}\end{figure}

\clearpage

\end{document}